*Original Research Article*

# Reliability of characterising coronary artery flow with the flow-split outflow strategy: comparison against the multiscale approach

## Authors:

Mingzi Zhang[*, †], Hamed Keramati[*], Ramtin Gharleghi, and Susann Beier

## Affiliations:

Sydney Vascular Modelling Group, School of Mechanical and Manufacturing Engineering, University of New South Wales, Sydney, NSW 2052, Australia.

[*]These authors contributed equally.

[†]Correspondence
Dr. Mingzi Zhang
E-mail: mingzi.zhang@unsw.edu.au
Telephone: +61 478 899 561

## Figure and Table counts:

Figure: 6 | Table: 2 | Supplementary Materials: 3

# Abstract

**BACKGROUND:**

In computational modelling of coronary haemodynamics, imposing patient-specific flow conditions is paramount, yet often impractical due to resource and time constraints, limiting the ability to perform a large number of simulations particularly for diseased cases.

**OBJECTIVE:**

To compare coronary haemodynamics quantified using a simplified flow-split strategy with varying exponents against the clinically verified but computationally intensive multiscale simulations under both resting and hyperaemic conditions in arteries with varying degrees of stenosis.

**METHODS:**

Six patient-specific left coronary artery trees were segmented and reconstructed, including three with severe (>70%) and three with mild (<50%) focal stenoses. Simulations were performed for the entire coronary tree to account for the flow-limiting effects from epicardial artery stenoses. Both a 0D-3D coupled multiscale model and a flow-split approach with four different exponents (2.0, 2.27, 2.33, and 3.0) were used. The resulting prominent haemodynamic metrics were statistically compared between the two methods.

**RESULTS:**

Flow-split and multiscale simulations did not significantly differ under resting conditions regardless of the stenosis severity. However, under hyperaemic conditions, the flow-split method significantly overestimated the time-averaged wall shear stress by up to 16.8 Pa ($p$=0.031) and underestimate the fractional flow reserve by 0.327 ($p$=0.043), with larger discrepancies observed in severe stenoses than in mild ones. Varying the exponent from 2.0 to 3.0 within the flow-split methods did not significantly affect the haemodynamic results ($p$>0.141).

**CONCLUSIONS:**

Flow-split strategies with exponents between 2.0 and 3.0 are appropriate for modelling stenosed coronaries under resting conditions. Multiscale simulations are recommended for accurate modelling of hyperaemic conditions, especially in severely stenosed arteries.

(247/250 words)

# 1. Introduction

Coronary Artery Diseases (CAD) are the leading cause of death worldwide, manifested primarily as atherosclerotic plaque buildup in major epicardial arteries. This buildup can result in stenoses or plaque rupture, obstructing the blood supply to heart tissues [1]. Over the past decade, patient-specific computational modelling of coronary blood flow has become an effective and clinically important approach to aid in the diagnosis and treatment of CAD [2,3]. To do this, Computed Tomographic Coronary Angiogram (CTCA) has been a critical modality due to its non-invasive nature and ability to simultaneously capture coronary anatomy along with fundamental stenosis characteristics.

CTCA-based coronary artery flow modelling facilitates the identification of flow-limiting stenoses [4] and has shown potential in identifying vulnerable plaques [5]. Amongst many blood flow metrics, CTCA-based Fractional Flow Reserve (FFR) is arguably the most important due to its widespread clinical use in assessing the functional significance of a stenosis, which is key to determining treatment strategies [6,7]. Other important blood flow metrics not yet clinically implemented include the Wall Shear Stress (WSS) [8,9], Oscillatory Shear Index (OSI) [10], and Relative Residence Time (RRT) [11] measured within the stenosed segments. Their diagnostic or prognostic value has been extensively demonstrated in research settings, marking them as drivers of plaque initiation and progression [5,11–14]. However, accurately characterising these metrics requires extensive computational expertise, significant resources, and time. Additionally, defining patient-specific boundary conditions for these simulations often necessitates invasive and costly methods, such as intravascular ultrasound, intracoronary continuous thermodilution, or related techniques [15].

Recent advances in computational physiology have demonstrated the reliability of 0D-3D coupled multiscale models in reproducing physiologically accurate coronary haemodynamics without invasive measurements [2,16–18]. These models employ a lumped parameter network (0D) to provide dynamic and personalised boundary conditions for the fully resolved 3D blood flow fields [16,19], enabling accurate derivation of FFR and other metrics [7]. However, such multiscale approaches require great computational cost and are highly complex, requiring expert adjustments for the lumped parameters for each patient. Moreover, these models require data exchange at every timestep between the lumped model and the fully resolved 3D model, imposing a greater difficulty for the multiscale simulation to reach global convergence.

As an alternative, defining a blood flow split ratio at bifurcations based on various scaling laws has been recommended as a more time- and resource-efficient outflow strategy [15]. Stemming from the form-follows-function concept in circulatory systems [20], flow-split ratios can be simply derived from the diameters of the daughter branches commonly referred to as the Murray's Law [21]. For coronary artery flow, specific exponents relating the daughter branch diameters to flowrates have been proposed, ranging from 2.0 to 3.0 [22–25]: (1) $k$ = 2.0, derived by assuming least energy expenditure required for the conduction of fluid through the entire coronary artery tree [25], (2) $k$ = 2.27, based on a least-square fitting of *in vivo* blood flow measurements [23,24], (3) $k$ = 2.33, following the minimum energy hypothesis customised for turbulent flows modelling [20], and (4) $k$ = 3.0, the universal exponent in Murray's Law invariant for all vascular trees with internal flows obeying laminar conditions [20,21].

Despite these developments, no previous work has statistically compared haemodynamic metrics as a result of the different flow-splitting exponents, even though low sensitivity of TAWSS to flowrate variations under the resting condition is reported [26]. Furthermore, flow-split methods with different scaling exponents have not been directly compared against *in vivo* measurement or multiscale simulations, especially considering their limitations in accounting for stenosis-induced flow redistributions within epicardial arteries [17,27]. These gaps restrict the clinical applicability of flow-split methods for rapid clinical assessments of coronary haemodynamics.

Herein, we assumed the 0D-3D coupled multiscale simulation to be the reference standard for coronary haemodynamics quantification, because its strong agreement with direct and invasive clinical measurements has previously been demonstrated [28]. We performed the 0D-3D coupled multiscale simulation on six patient-specific left coronary artery trees and compared them with the flow-split strategies using four different scaling exponents from the literature. Through the comparison, we aimed to (1) quantify the differences of flow-split outflow strategy versus reference-standard 0D-3D coupled multiscale simulation, and to (2) evaluate the comparative effect of flow-split exponent under resting and hyperaemic flow conditions in arteries with different stenosis severities. We hypothesise that this will reveal the appropriateness of using simplified flow-split strategy for modelling and clarify the uncertainties surrounding different scaling exponents in coronary blood flow modelling, which may ultimately inform the optimisation of clinical diagnostic approaches in the future.

## 2. Methods

### 2.1 Patient selection and coronary model preparation

As the flow-split method mostly differs from multiscale modelling in characterising severely stenosed arteries, three left coronary artery trees with severe stenoses (percent Diameter Stenosis [%DS] > 70) were randomly selected from the ASOCA [29] challenge (https://reshare.ukdataservice.ac.uk/855916/). These were compared to three arteries with mild stenoses (%DS ≤ 50) from the same dataset. **Table 1** presents the patients' demographics, locations and degrees of the stenoses. A diagram of patient selection and study design is shown in **Figure 1**. Access to the patients' data was approved by the institutional ethics committees of the University of New South Wales (HC190145) and the University of Auckland (022961).

Acquisition and reconstruction is described in detail in previous works [29,30]. Briefly, the stenosed coronary arteries were reconstructed from CTCA images obtained using a GE LightSpeed 64 slice CT scanner with an ECG-gated retrospective acquisition protocol. All images were annotated manually and independently for coronary lumen boundaries by three experts, before a majority voting method was applied resulting in a high-fidelity segmentation quality [29]. Due to the resolution of CTCA and lack of clinical relevance, distal coronary branches were clipped if < 2 mm in diameter. As per previous work [14], side branches without stenoses and with diameters smaller than one-third of the main vessels were removed using the Vascular Modelling ToolKit (VMTK, version 1.4) [31]. The length of a stenosed segment was defined as twice the diameter of the adjacent non-stenosed vessel, upstream and downstream of a focal stenosis.

## 2.2 Flow-split outflow strategies

Various scaling laws were established to split coronary blood flow at bifurcations per the Diameters ($D$) of daughter branches with an exponent ($k$) ranging between 2.0 and 3.0:

$$Q_i = \frac{D_i^k}{\sum_{m=1}^{n} D_m^k} Q_{\text{inflow}} \qquad (1)$$

where $n$ is the number of daughter branches at a bifurcation, $Q_i$ refers to blood flowrate at the $i^{th}$ daughter branch, $Q_{\text{inflow}}$ is the incoming blood flowrate at the proximal main vessel, and $D$ is the average diameter over the length of the daughter branch until a downstream bifurcation is reached. Here, we tested four different exponents ($k$ = 2.0, $k$ = 2.27, $k$ = 2.33, and $k$ = 3.0) as recommended in previous literature [22–25].

For the inflow condition, we adopted an uniform velocity profile, as recommended for modelling blood flow at the left main ostium [15], and a standard waveform from the literature [32], after scaling according to the inlet diameter $D$ of each patient, a common approach in literature after a strong agreement ($r^2$=0.87) with intravascular Doppler measurements was demonstrated [24]. This resulted in the scaled cycle-averaged flowrates $Q$:

$$Q = 1.43 D^{2.55}. \qquad (2)$$

To model the elevated or hyperaemic blood flow demand, we assumed the patient-specific cycle averaged flowrate $Q$ to be four times that of the resting condition in line with the literature [33]. The vascular geometries and the diameters calculated for each branch for scaling the inflow are presented in **Supplementary Material 1**.

## 2.3 Lumped parameter models

Patient-specific 0D-3D coupled multiscale models were established for the coronary tree to account for the flow-limiting effects of stenoses. This model is considered the reference standard and has been verified against direct invasive measurements for quantifying pressure drops across the coronary artery tree [28]. In the multiscale model, lumped parameter networks (0D) were coupled to the distal ends of the 3D coronary artery tree to model the resistance and compliance of the arterioles and capillary vessel beds, as well as the ventricular pressure reflecting coronary physiology [18,34]. At a 0D–3D interface (**Figure 2**), blood pressures ($P$), or flowrates ($Q$) were exchanged following [32]:

$$a \frac{d^2 P}{dt^2} + b \frac{dP}{dt} + P = c \frac{d^2 Q}{dt^2} + d \frac{dQ}{dt} + eQ + f \qquad (3)$$

where the coefficients *a* to *g* were calculated as:

$$a = R_v C_{im} R_{a-m} C_a \tag{4}$$

$$b = R_v C_{im} + R_v C_a + R_{a-m} C_a \tag{5}$$

$$c = R_{a-m} R_a R_v C_{im} C_a \tag{6}$$

$$d = R_a R_v C_{im} + R_{a-m} R_v C_{im} + R_a R_v C_a + R_{a-m} R_a C_a \tag{7}$$

$$e = R_a + R_{a-m} + R_v, \text{and} \tag{8}$$

$$f = R_v C_{im} \frac{dP_{LV}}{dt}. \tag{9}$$

$R_a$, $R_{a-m}$, and $R_v$ denote resistances of the arterioles, arterial, and venous microcirculations, respectively. Their summation was calculated based on the area of each outlet after assuming a mean arterial pressure to be 100 mmHg, venous pressure 15 mmHg, and $\frac{R_a}{R_{a-m}+R_a}$ = 0.38 [34]. The total compliance for the whole left coronary tree was adopted from the literature [34,35] and split across the branches proportional to the cross-section areas. Following previous studies, the relationship between the arterial ($C_a$) and myocardial ($C_{im}$) compliance was assumed to be $\frac{C_a}{C_{im}+C_a}$ = 11% [36]. This lumped parameter approach has previously been demonstrated to reproduce realistic coronary artery flow [36].

## 2.4 Computational fluid dynamics

Tetrahedron-predominant computational grids of each left coronary artery tree were generated using ICEM-CFD (ANSYS, version 2023R1, Canonsburg, USA), with the maximum sizes of the surface and volume elements being 0.1 and 0.2 mm. Five prismatic layers adhering to the arterial wall were generated with an increase ratio of 1.2. The total element numbers ranged between 1.4 and 3.3 million across different cases depending on their sizes. These values were deteremined based on a mesh sensitivity analysis [27].

As commonly applied in literature, we assumed the arterial wall to be rigid and assigned a no-slip condition [37]. The blood flow was modelled as an incompressible and non-Newtonian fluid, with the Carreau-model [38] employed to reflect the blood's shear-thinning behaviour:

$$\mu = \mu_i + \frac{\mu_0 - \mu_i}{[1 + (\lambda|\dot{\gamma}|)^b]^a} \tag{10}$$

where $\mu$ is the viscosity, $\mu_i$ is 0.0035 Pa·s as the high shear viscosity, $\mu_0$ is 0.16 Pa·s as the low shear viscosity, $\lambda$ is 8.2s as a time constant, and *a* and *b* are fixed values, *i.e.*, 0.64 and 1.23, respectively.

Transient coronary haemodynamics was resolved using ANSYS-Fluent CFD solver (version 2023R1) for four cardiac cycles, with results extracted from the last cycle to minimise the transient start-up effects. To specify the coupled boundary conditions calculated from the lumped parameter model, the user-defined functions with ANSYS-Fluent were employed. A time-step of 0.001 second was used for the implicit 2[nd] order temporal discretisation scheme following a previous work [39], and the convergence criteria for each time step was determined as $10^{-4}$ for the normalised velocity and pressure.

## 2.5 Haemodynamic metrics

We quantified clinically relevant haemodynamic metrics respectively under the resting and hyperaemic conditions. They are the Time-Averaged Wall Shear Stress (TAWSS) for its correlation with plaque progression and vulnerability [40], Oscillatory Shear Index (OSI), and Relative Residence Time (RRT) for their correlations with plaque initiation and development [3,8,10,11], calculated as

$$\text{TAWSS} = \frac{1}{T}\int_0^T |\tau_\omega| dt \tag{11}$$

$$\text{OSI} = \frac{1}{2}\left(1 - \frac{\left|\int_0^T \vec{\tau_\omega} dt\right|}{\int_0^T |\vec{\tau_\omega}| dt}\right) \tag{12}$$

$$\text{RRT} = \frac{1}{(1 - 2 \times \text{OSI}) \times \text{TAWSS}} \tag{13}$$

where $\tau_w$ is the flow-induced shear stress vector at the luminal wall, and $T$ denotes the cardiac cycle period. Even though we did calculate OSI values here, the resulting absolute values were small throughout the cycle and across all cases (<0.1). Due to the inherent uncertainty of all computational simulations, we did not present those results within the main text but only in the **Supplementary Material 2**. Due to uncertainties around the cut-offs for determining adverse local haemodynamic environments [14], we reported only the spatially averaged values of the TAWSS, OSI, and RRT for comparison between different boundary condition settings.

Under the resting conditions, we calculated the instantaneous wave-free ratio (iFR), defined as the ratio between distal ($P_{\text{d-wf}}$) and proximal coronary pressure ($P_{\text{a-wf}}$) during the diastolic wave-free period [41]:

$$\text{iFR} = \frac{P_{\text{d-wf}}}{P_{\text{a-wf}}}. \tag{14}$$

Under the hyperaemic condition, *i.e.* coronary flowrate being four times that of the resting condition, we calculated the fractional flow reserve (FFR) as per clinical standard:

$$\text{FFR} = \frac{P_{\text{d}}}{P_{\text{a}}} \tag{15}$$

where $P_{\text{d}}$ is the pressure measured one diameter distal to a stenosis, and $P_{\text{a}}$ is the pressure at the left main coronary ostium.

## 2.6 Statistical analyses

The statistical analyses were performed using the R language-based software JASP (version 0.19). We considered haemodynamic metrics calculated from the 0D-3D coupled multiscale model to be the reference standard, since the accuracy in its derived FFR values compared to direct and invasive measures has previously been demonstrated [28]. This accuracy ensures reliable quantification of shear stress, given the primarily pressure-driven nature of coronary blood flow.

We first compared the flow-split using four different exponents respectively against the reference standard by paired-sample two-sided *t*-test or Wilcoxon signed rank tests, depending on the normality of samples confirmed using Shapiro-Wilk tests. We also compared the blood flow metrics calculated within the flow-split strategies for variances between different exponent groups by repeated measures ANOVA, which reveals the sensitivity in scaling exponent choice. Finally, to explore the effects of different splitting ratios on the haemodynamic assessment of stenoses at different severities, we calculated the Maximal Differences (MD) and Standard Errors (SE) in each of the calculated haemodynamic parameter, respectively for the mild and severe stenosis subgroup. Continuous variables were expressed as mean and Standard Deviation (SD), and categorical variables were given as counts and percentages. Throughout the study, a $p < 0.05$ was considered statistically significant.

## 3. Results

### 3.1 Resting coronary haemodynamics via multiscale versus flow-split approach

When comparing the reference-standard multiscale model with flow-split approach under resting conditions using a paired-sample comparison, no significant differences were found between the two approaches regardless of the flow-split exponent used (TAWSS: $p > 0.233$, RRT: $p > 0.282$, and iFR: $p > 0.227$).

The flow-split strategy slightly overestimated TAWSS and underestimated RRT and iFR (**Figure 3**). Resting TAWSS decreased as the exponent ($k$) increased ($k$ = 2.0: 2.85 ± 2.83 Pa *vs.* $k$ = 3.0: 2.30 ± 2.16 Pa), with the highest exponent ($k$ = 3.0) most closely matching the multiscale simulation (2.46 ± 2.20 Pa). Resting RRT increased from 1.83 ± 1.05 Pa$^{-1}$ at $k$ = 2.0 to 2.51 ± 1.84 Pa$^{-1}$ at $k$ = 3.0, with 2.02 ± 1.08 Pa$^{-1}$ at $k$ = 2.33 best aligning with the multiscale results (2.15 ± 1.51 Pa$^{-1}$). Like TAWSS, iFR increased from 0.91 ± 0.16 at $k$ = 2 to 0.94 ± 0.10 at $k$ = 3.0, with the largest exponent ($k$ = 3.0) producing results closest to the multiscale simulation (0.97 ± 0.06). This suggests that the choice of exponent may have a subtle influence on the estimated haemodynamic metrics. (**Supplementary Material 3**)

### 3.2 Hyperaemic coronary haemodynamics via multiscale versus flow-split approach

Under hyperaemia, statistically significant differences were observed, with TAWSS overestimated ($p = 0.031$) and FFR significantly underestimated ($p = 0.043$) by the flow-split approach. While there was a clear trend for the flow-split approach to underestimate RRT (**Figure 4**), this difference was not statistically significant ($p > 0.094$).

Interestingly, the flow-split TAWSS significantly decreased as the exponent ($k$) increased, from 24.33 ± 25.81 Pa at $k$ = 2.0 to 19.06 ± 19.08 Pa at $k$ = 3.0, increasingly agreeing with the multiscale solutions. Whilst the highest exponent, as in the resting conditions, provided results closest to the multiscale reference, the output remained higher than the multiscale simulation (15.6 ± 10.2 Pa), although not statistically significant anymore ($p = 0.219$).

Detailed statistics of all haemodynamic metrics investigated, including the mean and standard deviations, are provided in **Supplementary Material 3**.

FFR was significantly underestimated by the flow-split models, increasing from 0.63 ± 0.38 at *k* = 2.0 to 0.71 ± 0.31 at *k* = 3.0. Again, whilst the highest exponent (*k* = 3.0) approached the multiscale target of 0.81 ± 0.21, the value remained significantly underestimated ($p$ = 0.043) (**Figure 4**).

### 3.3 Haemodynamic comparison of different exponents for the flow-split approach

Repeated measures ANOVA revealed no statistically significant differences in flow-split methods with different exponents (*i.e. k* = 2.0, 2.27, 2.33 and 3.0) across all haemodynamic metrics (*i.e.*, TAWSS, RRT, or iFR/FFR, all $p$ > 0.141) for both resting and hyperaemic conditions (**Figures 3** and **4**).

### 3.4 Impact of stenosis severity for multiscale versus flow-split modelling approach

The effects of different modelling approaches varied between mild and severe stenoses, with notable disparities observed primarily for severe stenoses under hyperaemic conditions. It should be noted here that due to the small sample size, i.e. three mild and three severely stenosed cases, results presented as follows are indicative as further discussed in the limitation section of the discussion.

Under resting conditions, both mild (**Figure 5** left panel) and severe stenoses (**Figure 6**, left panel) were minimally affected by the modelling approach used, with only minor differences in TAWSS, RRT, or iFR across both categories. Compared to the reference multiscale modelling, the maximum MDs in TAWSS, RRT, and iFR were 1.17 Pa, 1.14 Pa$^{-1}$, and 0.12, respectively for severe stenoses. For mild stenoses, the disparities were negligible, with MDs in TAWSS, RRT, and iFR being only 0.11 Pa, 0.51 Pa$^{-1}$, and 0.01, respectively (**Table 2**).

Under hyperaemic conditions, for mild stenoses (**Figure 5**, right panel) the effect of the flow-split method was small, with the MDs in TAWSS, RRT, and FFR being only 1.23 Pa, 0.49 Pa$^{-1}$, and 0.02, respectively (**Table 2**). However, for severely stenosed cases (**Figure 6**, right panel), the results were notably affected by the modelling approach, with MDs in TAWSS, RRT, and FFR being 16.80 Pa, 0.74 Pa$^{-1}$, and 0.33 for the flow-split compared to the reference multiscale method.

For example, case 4 with severe stenosis (%DS > 70) at the 2$^{nd}$ obtuse diagonal branch, flow-split simulations showed a pronounced local concentration of high TAWSS within the stenosed region (**Figure 6**, red circle), whereby as the multiscale simulation only showed a mild concentration of high TAWSS in the same region. Additionally, high RRT concentrations distal to the stenosis were captured by the multiscale simulations but not in the flow-split simulation (**Figure 6**, black circle). This underscores the importance of using a multiscale model to accurately characterise hyperaemic flow in severely stenosed arteries.

## 4. Discussion

We compared the haemodynamic outcomes of flow-split simulations with varying exponents to multiscale simulations of left coronary trees with different degrees of stenosis under both resting and hyperaemic conditions.

Under resting conditions, the haemodynamics were similar between the multiscale model (reference standard) and the flow-split strategy across all exponents (2.0 to 3.0). Blood flowrates are lower in the resting condition compared to hyperaemia, which explains the negligible differences in the resulting TAWSS and iFR across the boundary conditions tested. The similarity in high TAWSS concentrations around stenoses justifies the use of any exponent for modelling resting coronary blood flow, regardless the severity of stenosis. This is advantageous because simulations using the flow-split settings required notably less computational resources, time, and expertise compared to multiscale modelling approaches as the current reference standard, which markedly improves the simulation efficiency.

This finding aligns with the previous report of a low sensitivity of TAWSS to flowrate variations under the resting condition [26]. Although prescribing a fixed amount of blood flow at each branch resulted in a slightly smaller iFR by the flow-split compared to the multiscale modelling, the difference was less than 5% and the results did not hinder the diagnostic accuracy when the clinical cut-off 0.89 is used for iFR to determine the functional severity of a stenosis [42]. Such findings agree with previous studies [43,44] for mild stenoses. However, it should be noted that in cases of extreme severity, enforcing the flow rates at distal branches as per flow-split approach, may underestimate iFR, particularly when the value is within 0.86 and 0.93, *i.e.* the clinical 'grey zone' of stenosis significance [45], where the benefits of revascularisation over conservative medication therapy is unclear, and thus may misinform clinical decision making. Future studies would benefit from expanding our efforts by including a wider range of stenosed arteries to interrogate the validity of using the flow-split strategy for non-invasive iFR estimation, thereby directly contributing to treatment planning and clinical decision-making.

FFR, a clinical biomarker assessed under the hyperaemic condition, is highly sensitive to the prescribed boundary conditions, as demonstrated by the statistically significant differences observed in our study between the flow-split and multiscale simulations. Computational simulation for FFR assessment has garnered attention, with most studies relying upon multiscale approaches [46,47]. Using a 0.75 cut-off to determine the need for invasive intervention [48], our findings suggest that the flow-split strategy is unsuitable for severely diseased cases due to a significant underestimation of FFR. Coronary blood flowrates under hyperaemia also has prognostic implication [49], with lower hyperaemic flowrates correlating to target vessel failure and myocardial infraction. For this reason, using the flow-split strategy would likely result in erroneous prognostic assessment.

Various flow-split exponent, *i.e.* 2.0 [25], 2.27 [23,24], 2.33 [22] and 3.0 [20,21], have been proposed in the literature over the past years, often based on comparison with *in vivo* animal measurements from arteries without stenosis. However, no previous work investigated the statistical difference in haemodynamic quantities because of these exponents or stenosis degrees. A stenosis acts as a structural resistance to blood flow in the distal arteries. The diameter-based scaling law, which determines the flow-split ratio, simplifies the quantification of stenosis-induced structural resistance and tends to overestimate blood flow in severely stenosed branches. This occurs because it averages the restricted lumen diameter with non-stenosed segments when calculating the mean branch diameter for flow-splitting. Naturally, this simplification does not introduce significant errors in less stenosed arteries where structural resistance is inherently low, or in arteries with low-speed native flows where the stenosis-induced pressure drop is small. This agrees with our findings of minor differences in the three mildly stenosed coronary arteries we examined. However, in cases of severe stenosis or high blood flowrate (*e.g.*, under hyperaemia), this simplification which lacks an account of true local blood flow resistance tends to overestimate the blood flow passing through the stenosed branch. As a result, it is not surprising that our results showed that the local TAWSS is overestimated, while iFR or FFR for the stenoses is underestimated.

Thus, one potential improvement to current diameter-based scaling law is to use a larger exponent to account for the increased effects of resistance under hyperaemia, thereby reducing the blood flow into the stenosed daughter branches. Our observations support this: as the exponents used in the flow-split method increased from $k$ = 2.0 to $k$ = 3.0, while there was no statistically significant difference between them, the simulation results approached those of the multiscale simulation, particularly for severely stenosed coronaries under hyperaemia. If this adjustment can enhance accuracy, significant opportunities arise to simplify diagnostic virtual iFR modelling. Further studies are needed to refine the scaling exponents, incorporating the characteristics of stenosed vessel length, plaque eccentricity, and stenosis location.

In conclusion, although multiscale modelling remains the preferred approach for simulating severely stenosed coronaries under hyperaemia, discrepancies under resting conditions were not statistically significant, even for severely stenosed cases. This supports the use of the flow-split approach with an exponent of 3.0 across different levels of stenosis severity. Replacing multiscale simulations with the flow-split approach—using appropriate scaling exponents and standardised simulation protocols—would markedly reduce manual intervention while maintaining the simulation accuracy.

In our case, multiscale simulation required an average of six hours for an expert to derive and tune the lumped parameters, compared to the end-to-end automation of the flow-split outflow settings which require no manual intervention throughout the simulation process. Adopting the flow-split method would therefore improve computational efficiency and ultimately enable faster coronary haemodynamics assessment in clinical settings.

This study has limitations. First, our sample size is relatively small, without evenly distributed stenoses across the mild, moderate and severe categories, or beyond the left coronary artery trees. While stenosis-induced structural resistance is proportional to its degree, moderate stenoses, or stenoses in other segments of the coronary artery tree should still obey the trends found in this study, as shown in a previous study [27]. Simultaneously, the location and eccentricity of stenoses, vessel tortuosity, together with their impacts on flow-split ratio determination still warrant future investigations. Nevertheless, this study represents the first attempt to quantify the differences between flow-split strategies to multiscale simulations for mildly and severely stenosed coronary trees under both resting and hyperaemic conditions. Secondly, although we have used a non-Newtonian model, the uncertainties around different rheology models and the assumption of a rigid wall have not been tested. Still, assumptions made in this work are widely accepted in the literature [50,51]. Finally, we considered the previously verified multiscale simulation to be the reference standard, without matching the direct clinical measurements due to availability. However, previous work has repeatedly verified the pressure distribution across the coronary artery tree [28,52], and this verification has ensured reliable quantification of shear stress and its derivations, given the primarily pressure-driven nature of the coronary blood flow.

Overall, this work offers unique and useful insights into the suitability of flow-split approach compared to more complex, and resource intensive multiscale models directly informing diagnostic accuracy, clinical decision-making approaches and ultimately guiding optimisation of clinical treatment optimisation via FFR and iFR in future.

## Conclusions

Flow-split outflow conditions with exponents between 2.0 and 3.0 are suitable for modelling coronary artery flow under resting conditions, regardless of the stenosis severities, as they do not produce statistically significant differences compared to the reference-standard 0D-3D coupled multiscale modelling. However, for severe stenoses under hyperaemic conditions, flow-split strategy would significantly overestimate the TAWSS and underestimate the FFR, introducing potential uncertainties in the subsequent assessment of plaque vulnerability and functional severity, necessitating more computationally intensive multiscale simulations. Prevalent flow split exponent between 2.0 and 3.0 showed no difference in resulting haemodynamics, however higher exponent may be suitable for more stenosed cases.

## Figure Captions

**Figure 1.** Schematic of the study design from cases selection to simulation and analysis plans. We randomly selected three mildly and three severely stenosed left coronary artery trees and extracted clinically haemodynamic metrics from the multiscale simulations and simulations with the flow-split approach using four different scaling exponents, before analysing their differences respectively under resting and hyperaemic conditions.

**Figure 2.** Schematic of the 0D-3D coupled multiscale model for coronary artery flow in example Case 6 (middle), where a generic velocity waveform scaled per the inlet diameter (top left), same as that used in the flow-split outflow strategy, is prescribed at the left main inlet and 0D lumped parameter models representing the respective distal vascular beds are coupled to the 3D vessels' distal ends as outflow strategy (bottom right).

**Figure 3.** Resting Time-Averaged Wall Shear Stress (TAWSS, left), Relative Residence Time (RRT, middle), and instantaneous wave-Free Ratio (iFR, right) for the flow-split strategy using four different exponents (k = 2, 2.27, 2.33, and 3 left to right, black) compared against the multiscale strategy (green) as reference standard for three mild (cross) and three severe (triangle) stenoses cases. Results suggest no statistically significant difference between the multiscale and flow-split results using any of the exponents. (P-values are results of paired-sample *t*-tests or Wilcoxon tests between the multiscale simulation and the respective exponents used in flow-split strategy).

**Figure 4.** Hyperaemic Time-Averaged Wall Shear Stress (TAWSS, left), Relative Residence Time (RRT, middle), and instantaneous wave-Free Ratio (iFR, right) for the flow-split strategy using four different exponents (k = 2, 2.27, 2.33, and 3 left to right, black) against the multiscale strategy (green) as reference standard for three mild (cross) and three severe (triangle) stenoses cases. Results suggest that the hyperaemic TAWSS was significant overestimated and FFR significantly underestimated. (P-values are results of paired-sample *t*-tests or Wilcoxon tests between the multiscale simulation and the respective exponents used in flow-split strategy)

**Figure 5.** Time-Average Wall Shear Stress (TAWSS, top), Oscillatory Shear Index (OSI, middle), and Relative Residence Time (RRT, bottom) under resting (left panel) and hyperaemic (right panel) conditions computed via flow-split (sub-left) and multiscale simulations (sub-right) for an example coronary artery tree (Case 3) with mild stenosis located at the middle left anterior descending artery (black-dotted circle), which exhibits no marked difference under either the resting or hyperaemic condition.

**Figure 6.** Time-Average Wall Shear Stress (TAWSS, top), Oscillatory Shear Index (OSI, middle), and Relative Residence Time (RRT, bottom) under resting (left panel) and hyperaemic (right panel) conditions computed via flow-split (sub-left) and multiscale simulations (sub-right) for an example coronary artery tree (Case 4) with a severe stenosis located at the 1st diagonal branch, which exhibits noteworthy differences in TAWSS (red-dotted circle) and RRT (black-dotted circle) under hyperaemia but not under the resting condition.

## Declaration

Access to the patients' data was approved by the institutional ethics committees of the University of New South Wales (HC190145) and the University of Auckland (022961).

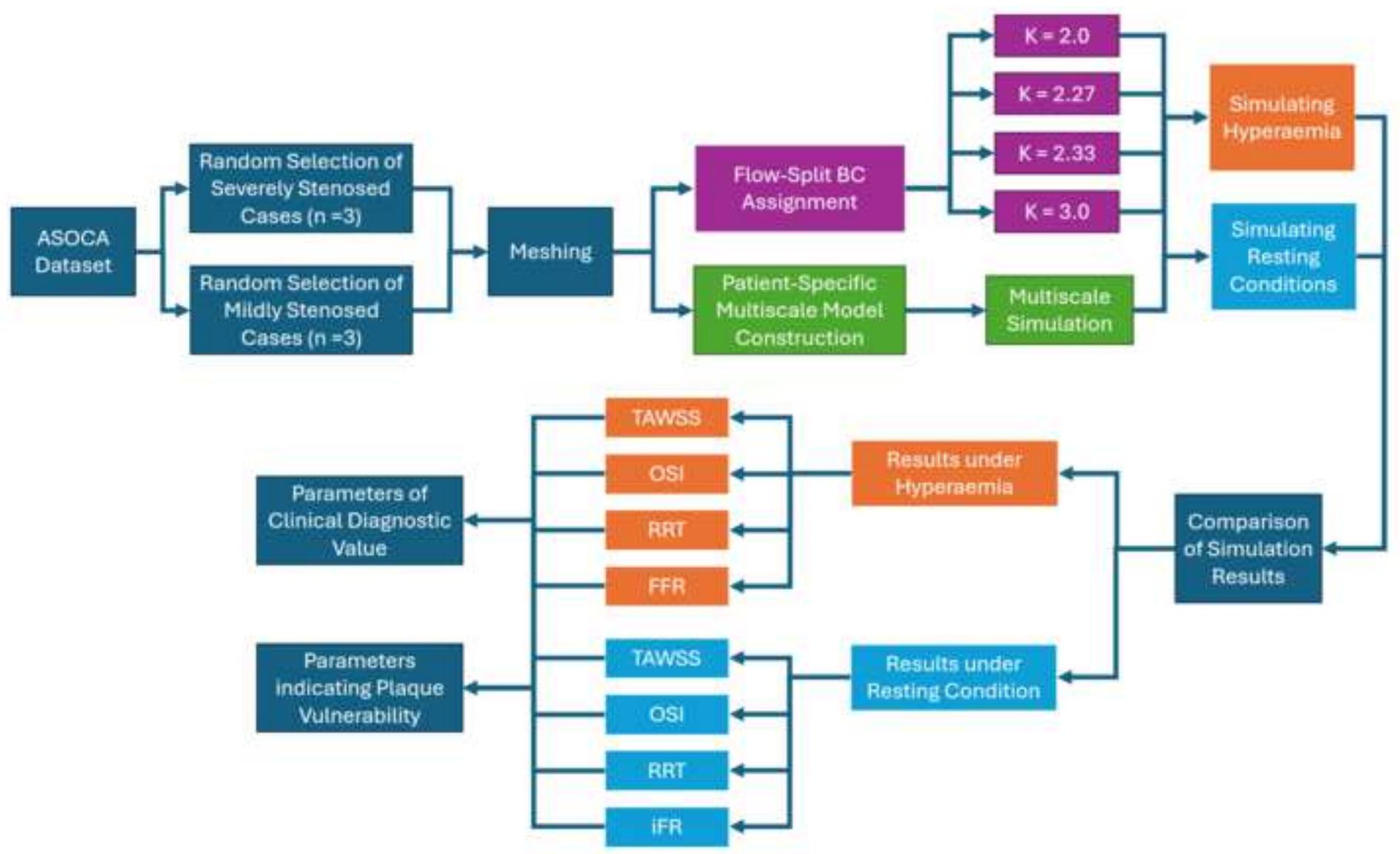

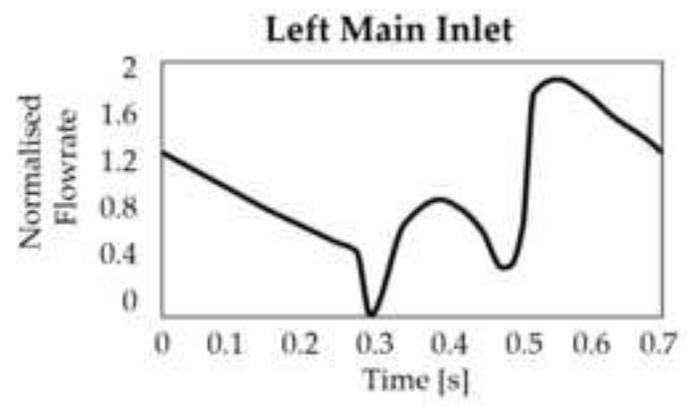
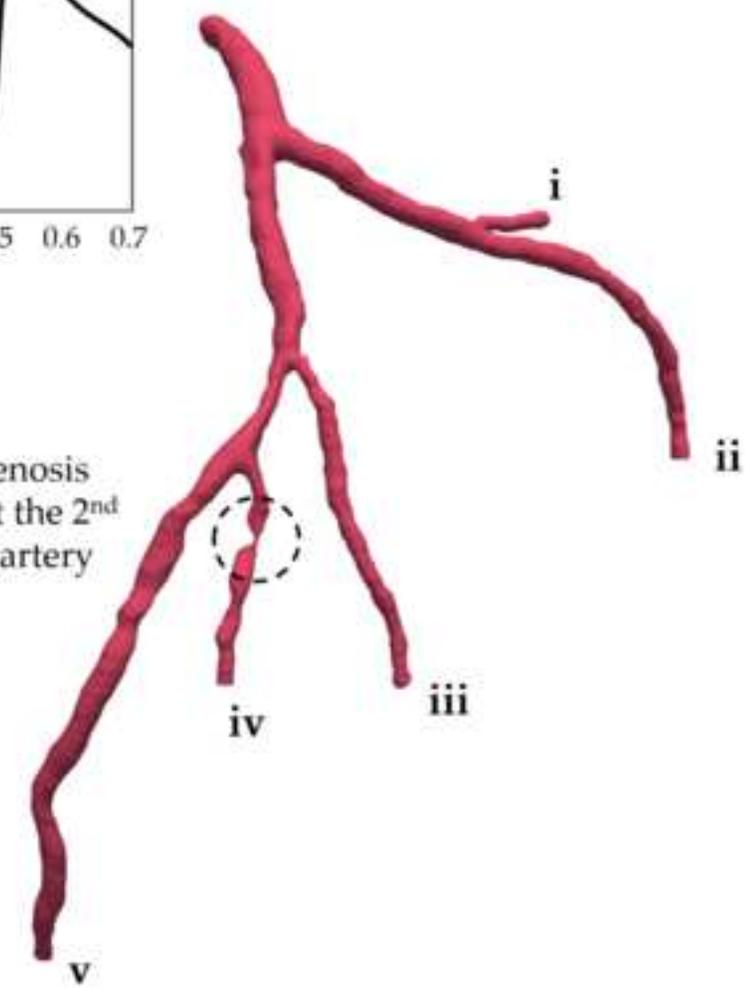
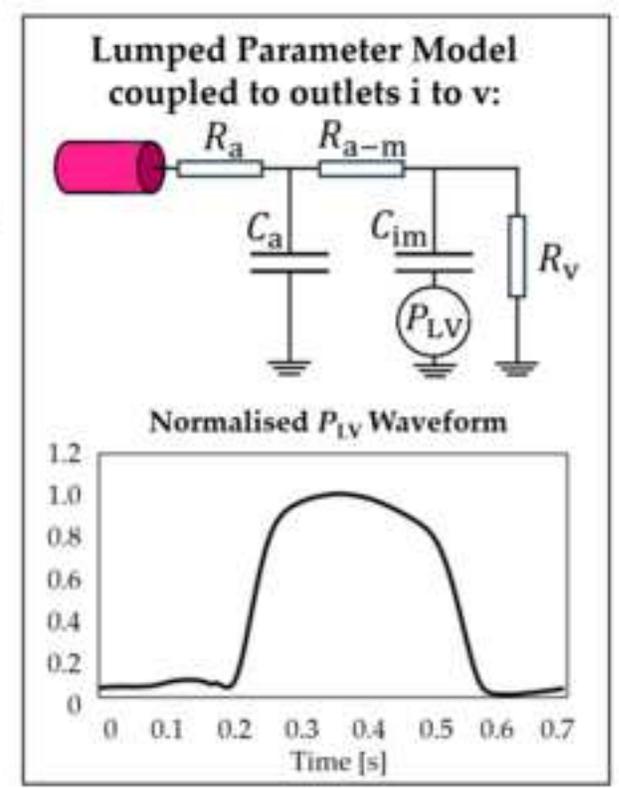

Fig_3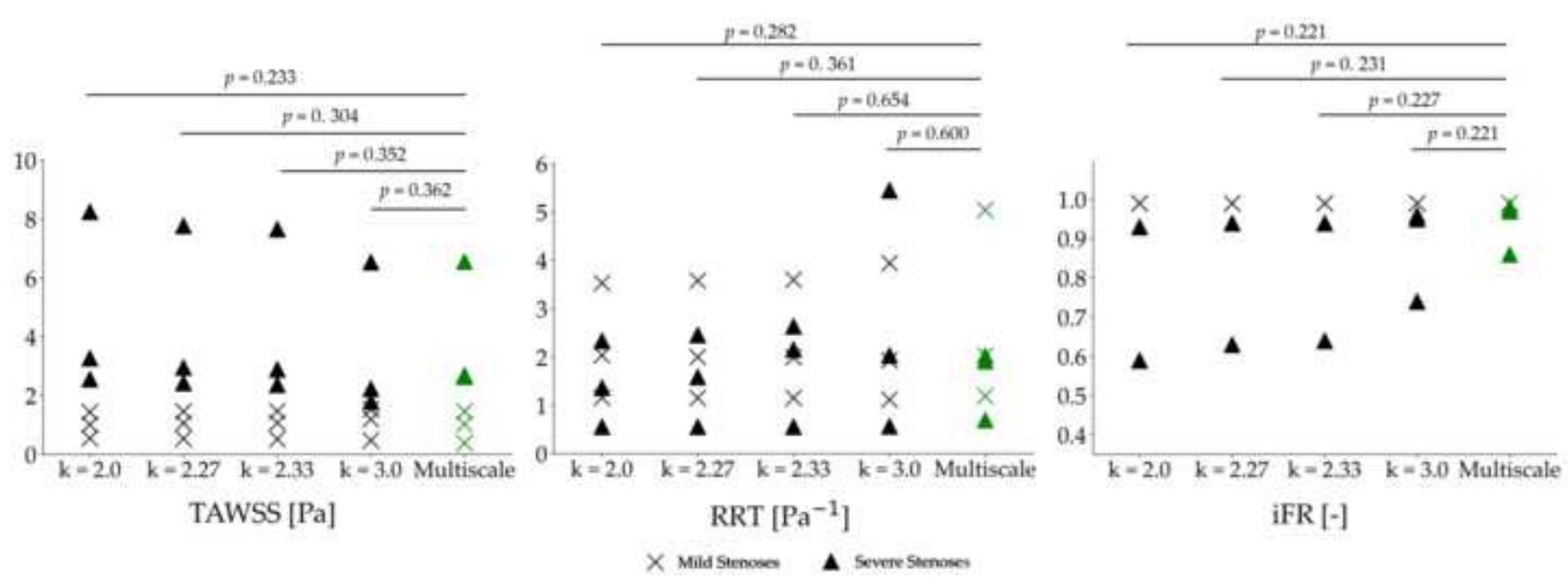

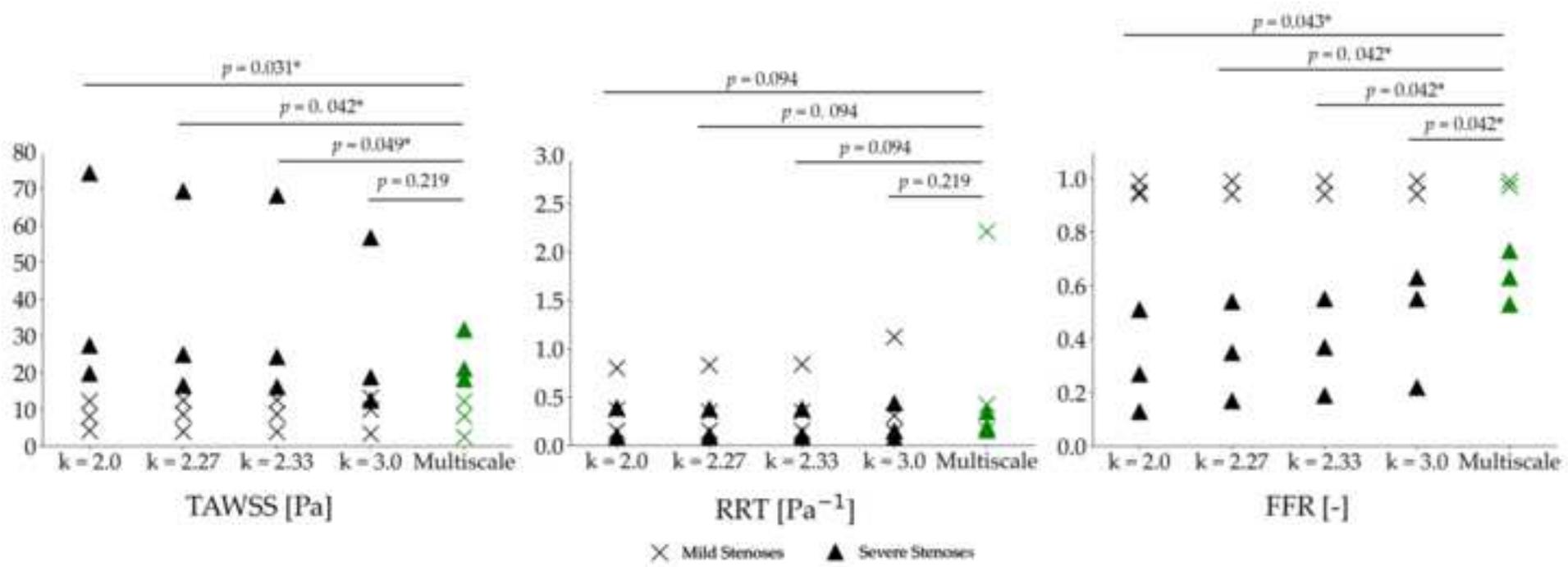

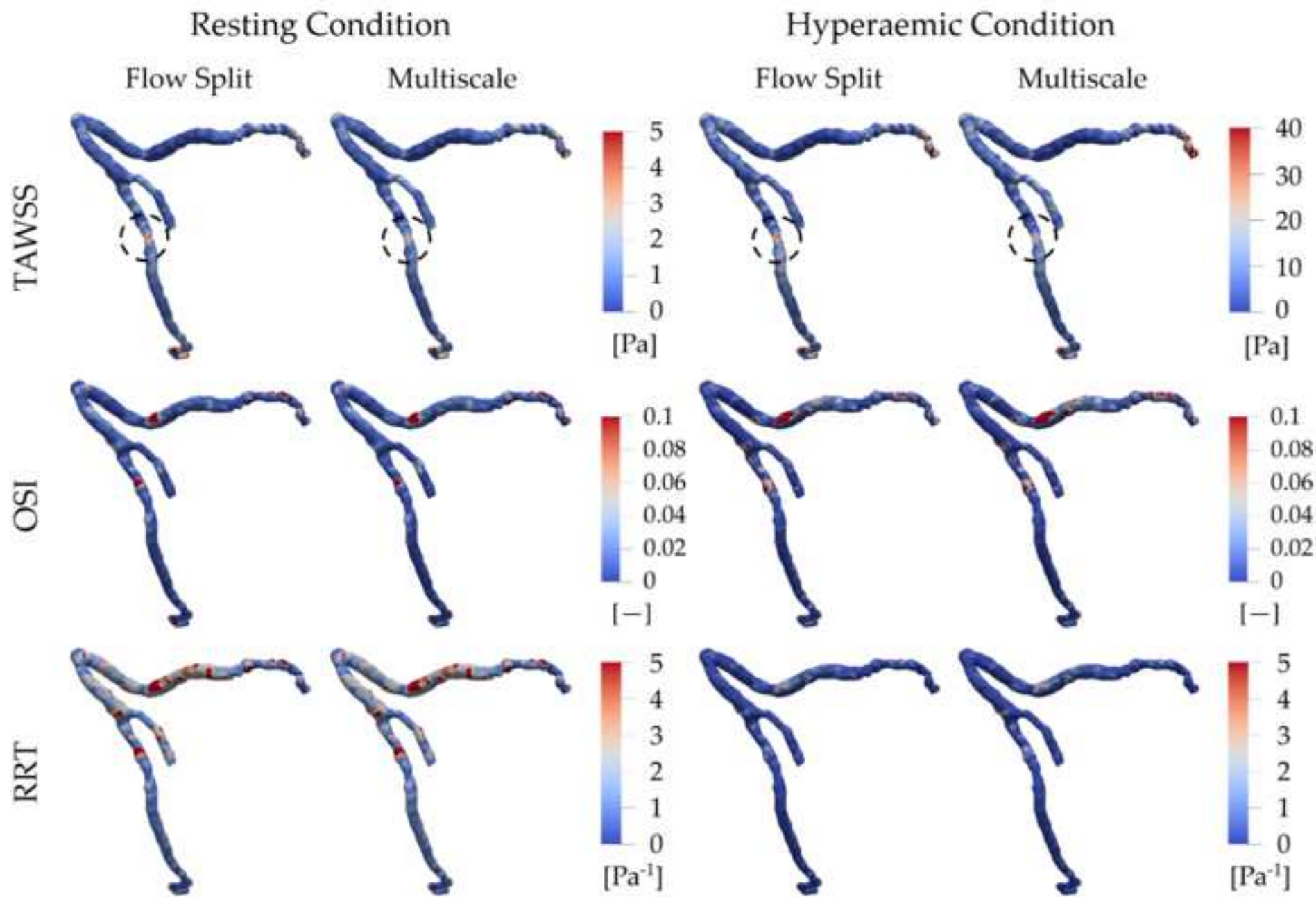

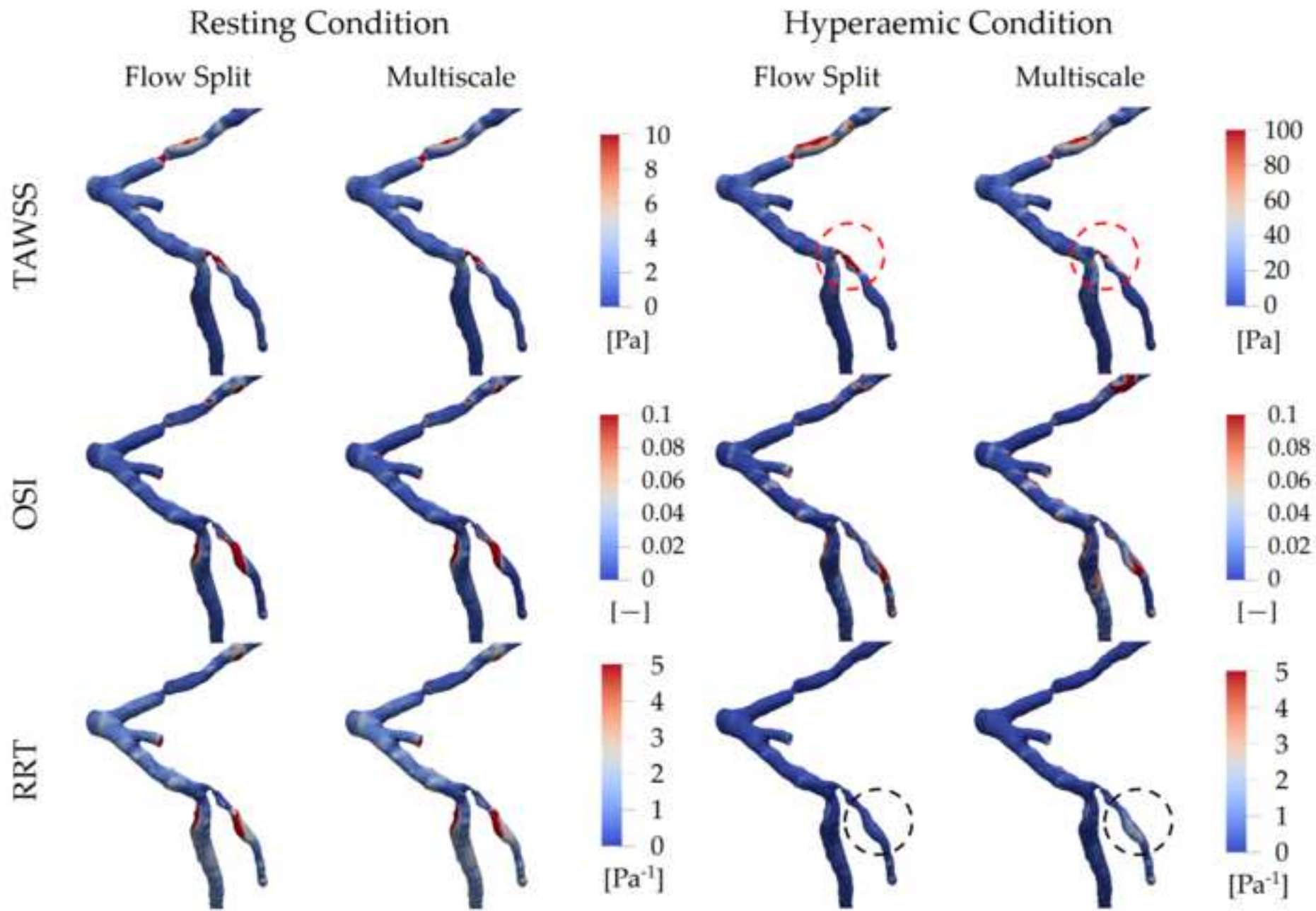

# Table 1

**Table 1.** Patient demographics and lesions characteristics randomly selected from the ASOCA dataset [29].

| Case | Sex | Age | Weight (kg) | Height (m) | BMI | SBP (mmHg) | DBP (mmHg) | Smoker | Hypertension | Stenosis Location | % Diameter Stenosis | CAD-RADS Score |
|---|---|---|---|---|---|---|---|---|---|---|---|---|
| \multicolumn{13}{Moderate stenosis} | | | | | | | | | | | | |
| 1 | F | 55 | 83 | 1.7 | 29 | 149 | 87 | P | Y | LAD | 25-49% | 2 |
| 2 | M | 64 | 60 | 1.64 | 22 | 179 | 82 | P | N | LAD | 25-49% | 2 |
| 3 | F | 55 | 67 | 1.67 | 24 | 110 | 66 | N | N | LAD | 1-24% | 1 |
| \multicolumn{13}{Severe stenosis} | | | | | | | | | | | | |
| 4 | M | 55 | 73 | 1.72 | 25 | 125 | 77 | N | N | $D_2$ | 70-99% | 4A |
| 5 | F | 78 | 104 | 1.79 | 32 | 128 | 91 | N | Y | $M_1$ | 70-99% | 4A |
| 6 | M | 54 | 77 | 1.8 | 24 | 107 | 67 | N | N | $D_2$ | 70-99% | 4A |

Note: LAD = Left Anterior Descending artery, $D_1$ = The first Diagonal artery, $D_2$ = The second Diagonal artery, $M_1$ = The first obtuse Marginal artery, SBP = Systolic Blood Pressure, DBP = Diastolic Blood Pressure, P or N for Smoker indicate past and non-smokers, Y or N for hypertension present and not respectively.

## Table 2

**Table 2.** Maximum differences in the Time-Averaged Wall Shear Stress (TAWSS), Oscillatory Shear Index (OSI), Relative Residence Time (RRT), Fractional Flow Reserve (FFR) for the hyperaemic condition, and Instantaneous wave-Free Ratio (iFR) for the resting condition within the stenosed region between the flow-split outflow strategy and 0D-3D coupled multiscale simulation.

|  |  | Resting | | Hyperaemic | |
| --- | --- | --- | --- | --- | --- |
|  | DS | Maximum Mean Differences | SE | Maximum Mean Differences | SE |
| TAWSS [Pa] | Mild | 0.11 | 0.18 | 1.23 | 6.87 |
|  | Severe | 1.17 | 0.21 | 16.80 | 9.15 |
| OSI [-] | Mild | 0.01 | 0.01 | 0.01 | 0.01 |
|  | Severe | 0.01 | 0.01 | 0.02 | 0.02 |
| RRT [Pa$^{-1}$] | Mild | 0.51 | 0.40 | 0.49 | 0.32 |
|  | Severe | 1.14 | 0.84 | 0.74 | 0.48 |
| FFR [-] | Mild | — | — | 0.02 | 0.03 |
|  | Severe | — | — | 0.33 | 0.04 |
| iFR [-] | Mild | 0.01 | 0.02 | — | — |
|  | Severe | 0.12 | 0.05 | — | — |

*Note: DS = Diameter Stenosis and SE = Standard Errors.*